\documentclass{PoS}

\title{Measurements of CKM angles at Belle}

\ShortTitle{CKM angles at Belle}

\author{\speaker{Bilas Pal}\thanks{On behalf of the Belle Collaboration}\\
        University of Cincinnati\\
        E-mail: \email{palbs@ucmail.uc.edu}}

\abstract{
In this review recent studies on $CP$ violation and related hadronic $B$ decays by the Belle experiment, in particular measurements of CKM angles $\phi_1$ and $\phi_2$ are reported.
  }

\FullConference{ 9th International Workshop on the CKM Unitarity Triangle\\
                 28  November - 3 December 2016\\
                 Tata Institute for Fundamental Research (TIFR), Mumbai, India}

\begin{document}

\section{Introduction}
In the standard model (SM) of electroweak interaction,  charge-parity $(CP)$ violation arises from an  irreducible complex phase
in the Cabibbo-Kobayashi-Maskawa (CKM) quark-mixing matrix~\cite{ckm}.  The Belle and BaBar experiments have established $CP$ violating effects  in the $B$ meson system. Both experiments use their measurements of the mixing-induced $CP$ violation in $b\to c\bar{c}s$ transitions~ to precisely determine the parameter $\sin (2\phi_1)$, where $\phi_1$ is defined as $\arg[-V_{cd}V^*_{cb}/V_{td}V^*_{tb}]$, with $V_{ij}$ is the CKM matrix element of quarks $i,~j$. In this proceeding an overview  of  recent measurements of the CKM angles  $\phi_1$ and $\phi_2$ $\big(\arg[-V_{td}V^*_{tb}/V_{ud}V^*_{ub}]\big)$ is presented. Unless stated otherwise, all measurements presented here  are based on Belle's final dataset of $772\times10^6$ $B\bar{B}$ pairs.

\section{First observation of $CP$ violation in $\bar{B}^0\to D_{CP}^{(*)}h^0$ decays with Belle + BaBar data}
The decay $\bar{B}^0\to D^{(*)}h^0$, where $h^0$ is a light, unflavored neutral meson ($h^0\in{\pi^0,~\eta,~\omega}$), is dominated by a $b\to c\bar{u}d$ color-suppressed tree diagram in the SM. The final state $D^{(*)}h^0$ is a $CP$ eigenstate if the neutral $D$ meson  decays to a $CP$ eigenstate as well ($i.e.$,  $D^0_{CP}\to K_S^0\pi^0$, $D^0_{CP}\to K_S^0\omega$ ($CP=-1$) or $D^0_{CP}\to K^+K^-$ ($CP=+1$) and $D^*_{CP}\to D^0_{CP}\pi^0$). Therefore, a time-dependent $CP$ asymmetry measurement is applicable in the same way as used in the $b\to c\bar{c}s$  decays, but with a small correction from the $b\to u\bar{c}d$ process. This $b\to u\bar{c}d$ amplitude is suppressed by $V_{ub}V^*_{cd}/V_{cb}V^*_{ud}\approx 0.02$ relative to the leading amplitude.
Neglecting the suppressed amplitude, the time evolution of $\bar{B}^0\to D_{CP}^{(*)}h^0$ decays is governed by $\phi_1$\cite{Fleischer:2003ai}.

\begin{figure}[htb]
\centering
\includegraphics[width=0.65\textwidth]{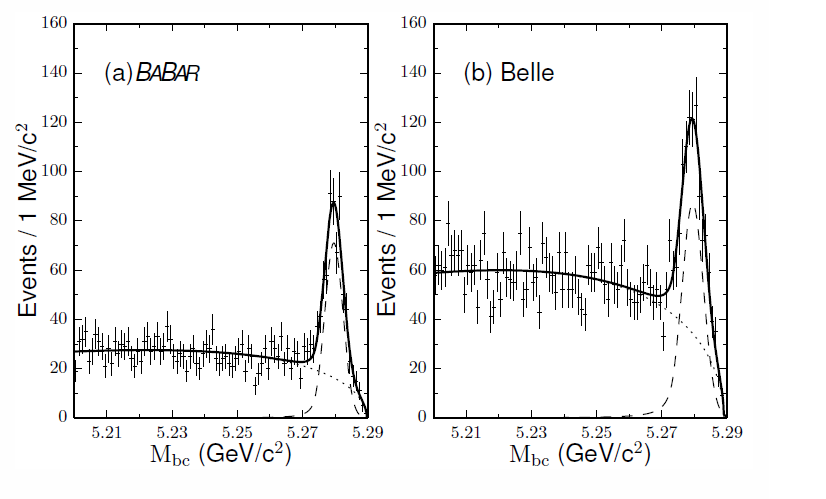}%
\vskip -0.6cm
\caption{\small $M_{\rm bc}$ distributions (data points with error bars) and fit projections (solid lines) of $\bar{B}^0\to D_{CP}^{(*)}h^0$ decays for (a) BaBar and (b) Belle.  The dashed (dotted) lines represent
projections of the signal (background) fit components.}
\label{fig1}
\end{figure}

Due to the limited available statistics, previous measurements performed separately by the BaBar and Belle collaborations  were not able to establish $CP$ violation in these or related decays~\cite{Krokovny:2006sv}.  This motivated a joint analysis using the combined full dataset of Belle and BaBar experiments~\cite{Abdesselam:2015gha}. Using a fit to the beam-energy constrained mass $M_{\rm bc}=\sqrt{E^2_{\rm beam}-p^2_{B}}$, where $E_{\rm beam}$ is the beam energy and $p_{B}$ is the reconstructed $B$ meson momentum in the center-of-mass system, we extract $508\pm31$ signal events in the BaBar data of 431 million $B\bar{B}$ events and $757\pm44$ signal events in the Belle data of 772 million $B\bar{B}$ events, as shown in Fig.~\ref{fig1}. The dominant source of  background originates from $e^+e^-\to p\bar{q}~(q\in {u,~d,~s,~c})$ continuum events. To suppress this
background, we use a multivariate analyzer based on a neural network. The neural network uses
the so-called event shape variables to discriminate continuum events, which tend to be jetlike, from
spherical $B\bar{B}$ events. 

The time-dependent $CP$ violation measurement is performed using established Belle and BaBar techniques for the vertex reconstruction, the flavor-tagging, and the modeling of $\Delta t$ resolution effects, where $\Delta t$ is the proper time interval between the decays of the two $B$ mesons produced in an $\Upsilon(4S)$ decay. Combined analysis is performed by maximizing a joint log-likelihood function
\begin{equation}
\ln \mathcal{L} = \sum_i\ln \mathcal{P}_i^{\rm Belle} +  \mathcal{P}_i^{\rm BaBar}.
\end{equation}
The experiment-dependent probality density function (PDF) $\mathcal{P}^{\rm Exp}$ is defined as
\begin{equation}
\mathcal{P}^{\rm ExP} = \sum_k f_k \int\left[P_k(\Delta t')R_k^{\rm Exp}(\Delta t- \Delta t')\right]d\Delta t',
\end{equation}
where the index $k$ represents the signal and background PDF components. The symbol $P_k$ denotes the PDF describing the proper time interval of the particular physical process and $R_k^{\rm Exp}$ refers to the corresponding resolution function. The fractions $f_k$ are evaluated on an event-by-event basis as a function of $M_{\rm bc}$. While the background model is determined from the $M_{\rm bc}$ sideband and hence is experiment-dependent, the signal model is expressed as
\begin{equation}
P_{\rm sig}(\Delta t,q)=\frac{1}{4\tau_{B^0}}e^{\frac{-|\Delta t|}{\tau_{B^0}}}\left[1+q\left(S \sin(\Delta m\Delta t)- A\cos (\Delta m\Delta t)\right)\right],
\end{equation}
where the $B^0$ meson lifetime is represented by $\tau_{B^0}$, $B^0-\bar{B}^0$ mixing frequency by $\Delta m$ and $q$ is event- and experiment-dependent tagging quality parameter. In the SM, the coefficients, $S=-\eta_f\sin (2\phi_1)$
and $A=0$, where $\eta_f$ is the $CP$ eigenvalue of the final state. $S$ and $A$ quantify mixing-induced and direct $CP$ violation, respectively. The combined fit gives
\begin{equation}
-\eta_f S=+0.66\pm0.10~{(\rm stat.)} \pm 0.06~{(\rm syst.)}, ~~A=-0.02\pm0.07~{(\rm stat.)} \pm0.03~{(\rm syst.)}.
\end{equation}
These results correspond to the first observation of $CP$ violation in $\bar{B}^0\to D_{CP}^{(*)}h^0$ decays with a significance of 5.4 standard deviations and are in agreement with the value of $\phi_1$ measured from $b\to c\bar{c}s$ transitions.
\section{Measurement of $\phi_1$ in $B^0\to\bar{D}^{(*)0}h^0$ with time-dependent binned Dalitz plot analysis}
\begin{figure}[htb]
\centering
\includegraphics[width=0.8\textwidth]{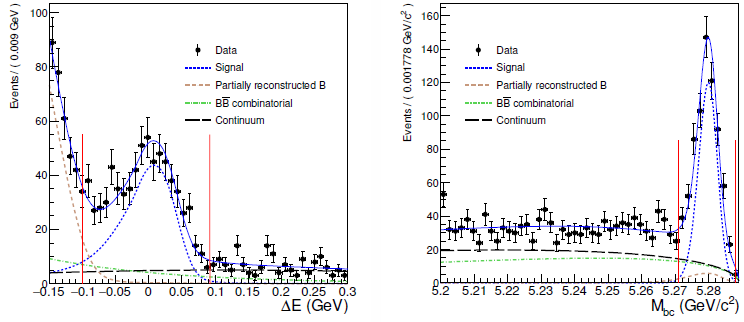}%
\vskip -0.4cm
\caption{\small $M_{\rm bc}$ and $\Delta E$  distributions of $B^0\to \bar{D}^0\pi^0$ decays.}
\label{fig2}
\end{figure}

In this analysis,   we present a model-independent measurement of the angle $\phi_1$ in $b\to c\bar{u}d$ transitions governing $B^0\to\bar{D}^{(*)0}h^0$ decays, with subsequent decay $\bar{D}^0\to K_S^0\pi^+\pi^-$ is not a $CP$ eigenstate~\cite{Vorobyev:2016npn}. From a fit to $M_{\rm bc}$ and $\Delta E= E_B-E_{\rm beam}$, where $E_B$ is the reconstructed $B$ mesons energy in the center-of-mass system, we extract total $962\pm41$ signal events, of which $464\pm26$ events are from $B^0\to\bar{D}^0\pi^0$ mode (Fig.~\ref{fig2}), with a signal fraction $(72.1\pm4.1)\%$ and $182\pm18$ events from $B^0\to\bar{D}^0\omega$ with a fraction of $(58.4\pm5.7)\%$. The signal fraction of other decay modes ranges between 44\% and 70\%.   

Our measurement of $\phi_1$ is based on the binned Dalitz distribution approach. This idea was proposed in Ref.~\cite{Giri:2003ty} to measure the angle $\phi_3$. Events are divided into 16 bins on the Dalitz plot plane and the number of events in bin $i~(i=-8, ..., -1, +1, ..., +8)$ is modeled as
\begin{eqnarray}
P_i(\Delta t,\phi_1)&=&he^{-\frac{|\Delta t|}{\tau_B}}\Big[1+q\frac{K_i-K_{-i}}{K_i+K_{-i}}\cos(\Delta m\Delta t) \nonumber\\
&+&2q\xi_{h^0}(-1)^L\frac{\sqrt{K_iK_{-i}}}{K_i+K_{-i}}\sin(\Delta m\Delta t)\big(S_i\cos(2\phi_1)+C_i\sin(2\phi_1)\big)\Big],
\end{eqnarray}
where  $h$ is the normalization constant, $\xi_{h^0}$ is the $CP$ eigenvalue of $h^0$ meson, $L$ is the relative angular momentum in the $D^{(*)0}h^0$ system, $K_i$ is the integrated squared amplitude, and $S_i$ and $C_i$ represent the weighted averages of the sine and cosine of the phase difference between $\bar{D}^0$ and $D^0$ decay amplitudes over the $i$th Dalitz plot bin. The parameters $K_i$ can be measured with a set of flavor-tagged neutral $D$ mesons such as $D^{*+}\to D^0\pi^+$ or $B^+\to\bar{D}^0\pi^+$ decays, by measuring signal yield in each Dalitz plot bin. The measurement of the phase parameters $S_i$ and $C_i$ is more complicated and can be done with coherent decays of $D^0\bar{D}^0$ pairs~\cite{Libby:2010nu}. We obtain
\begin{eqnarray}
\sin(2\phi_1)&=&0.43\pm0.27~{(\rm stat.)}\pm0.08~{(\rm syst.)} \nonumber \\
\cos(2\phi_1)&=&1.06\pm0.33~{(\rm stat.)}^{+0.21}_{-0.15}~{(\rm syst.)} \nonumber \\
\phi_1 &=& 11.7^{\circ}\pm7.8^{\circ}~{(\rm stat.)}\pm2.1^{\circ}~{(\rm syst.)}.
\end{eqnarray}
The value $\sin(2\phi_1)=0.691\pm0.017$ measured in $b\to c\bar{c}s$ transitions determines the absolute value of $\cos(2\phi_1)$, leading two possible solutions in the $0^{\circ}\le\phi_1\le180^{\circ}$ range. Our measurement is inconsistent with the negative solution, corresponding to the value $\phi_1=68.1^{\circ}$ at the level of 5.1 standard deviations, but in agreement with the positive solution, corresponding to the value $\phi_1=21.9^{\circ}$ at 1.3 standard deviations. 
\section{First observation of the decay $B^0\to\psi(2S)\pi^0$}
Although decays mediated via $b\to c\bar{c}s$  transitions allow us to access the $\phi_1$ at first order (tree), its value is prone to distortion from suppressed higher-order loop-induced (penguin) amplitudes containing different weak phases. The related  $b\to c\bar{c}d$ induced decays can be used to quantify the shift in $\phi_1$ caused by these loop contributions and may provide useful information about the penguin pollution~\cite{Ciuchini:2005mg}. Since the dominant $b\to c\bar{c}d$  tree amplitude is also suppressed, $B^0\to J/\psi\pi^0$ is the only mode measured so far, providing $\sin (2\phi_1) = 0.65\pm0.21~{(\rm stat.)}\pm0.05~{(\rm syst.)}$ by Belle~\cite{Lee:2007wd}, which is consistent with $\sin (2\phi_1)$ from $b\to c\bar{c}s$. The possible next mode,  $B^0\to\psi(2S)\pi^0$ was not observed previously. 

The decay mode  $B^0\to\psi(2S)\pi^0$ is reconstructed with $\psi(2S)\to \ell^+\ell^-~(\ell=e,~\mu)$ or $\psi(2S)\to J/\psi(\to \ell^+\ell^-)\pi^+\pi^-$~\cite{Chobanova:2015ssy}. The major background contribution originates from $b\to c\bar{c}q$ decays other than the signal. The background arises from $e^+e^-\to q\bar{q}~(q=u,~d,~s,~c)$ continuum events is not so problematic and is suppressed by applying a loose requirement on the ratio of second- to zeroth-order Fox-Wolfram moments. The signal is extracted from a fit to $M'_{\rm bc}$ and $\Delta E$, where $M'_{\rm bc}$ is the modified beam-constrained mass to take into account the worse energy resolution of $\pi^0$ than rest of the particles. The fit gives $85\pm12$ signal events with a significance of 7.2 standard deviations. The branching fraction is measured to be
\begin{equation}
\mathcal{B}(B^0\to\psi(2S)\pi^0)=[1.17\pm0.17~{(\rm stat.)}\pm0.08~{(\rm syst.)}]\times 10^{-5}.
\end{equation}
This measurement constitutes the first observation of this decay and it will contribute to the future time-dependent $CP$ asymmetry measurement of the $b\to c\bar{c}d$ process.
\section{Study of $B^0\to\rho^+\rho^-$ decays}
In order to access $\phi_2$, charmless decay modes that are mediated via $b\to u\bar{u}d$ transitions are necessary.
Examples are the decays $B\to\pi\pi,~\rho\pi,~\rho\rho$. At tree level, one expects $A=0$ and $S=\sin(2\phi_2)$. 
Possible penguin contributions can give rise of direct $CP$ violation, $A\neq0$ and also pollute the measurement of $\phi_2$, $S=\sqrt{1-A^2}\sin(2\phi_2^{\rm eff})$, where 
the observed $\phi_2^{\rm eff}=\phi_2-\Delta\phi_2$ is shifted
by $\Delta \phi_2$ due to different weak and strong phases from additional non-leading contributions.
This inconvience can be overcome by estimating $\Delta \phi_2$ using either an isospin analysis~\cite{Gronau:1990ka} or $SU(3)$ flavor symmetry~\cite{Beneke:2006rb}. In this analysis, we present a measurement of the branching fraction and the longitudinal polarization fraction of $B^0\to\rho^+\rho^-$ decays, as well as the time-dependent $CP$ violating parameters~\cite{Vanhoefer:2015ijw}.

\begin{figure}[htb]
\centering
\includegraphics[width=0.5\textwidth]{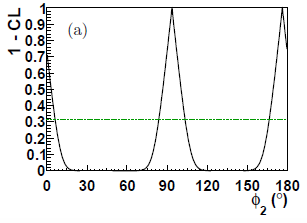}%
\vskip -0.6cm
\caption{\small Probability scan of $\phi_2$ in the $B\to\rho\rho$ system. The horizontal line shows the 68\% confidence level.}
\label{fig3}
\end{figure}

In addition to combinatorial background, the presence of multiple background components with the same four-pion final state as $B^0\to\rho^+\rho^-$ make this decay quite difficult to isolate  and interferences between the various four-pion modes need to be considered. A multi-dimensional maximum likelihood fit is performed. The fit uses the variables $\Delta E$, $M_{\rm bc}$, the masses and helicity angles (angle between one of the daughter of $\rho^{\pm}$ meson and the $B$ flight direction in the corresponding rest frame of the $\rho^{\pm}$.) of the two reconstructed $\rho^{\pm}$ mesons to
separate longitudinally polarized states from transversely polarized   states, a fisher discriminant to separate the jet-like continuum events from the spherical $B\bar{B}$ decays and the $\Delta t$ distribution for the two flavors of $B_{\rm tag}$.
We obtain the branching fraction
\begin{equation}
\mathcal{B}(B^0\to\rho^+\rho^-)=[28.3\pm1.5~(\rm stat.)\pm 1.5~(syst.)]\times10^{-6},
\end{equation}
the fraction of longitudinal polarization
\begin{equation}
f_L=0.988\pm0.012~(\rm stat.)\pm0.023~(syst.),
\end{equation}
and the $CP$ violating parameters
\begin{equation}
S=-0.13\pm0.15~{(\rm stat.)} \pm 0.05~{(\rm syst.)}, ~~A=0.00\pm0.10~{(\rm stat.)} \pm0.06~{(\rm syst.)}.
\end{equation}
These results together with the other Belle measurements~\cite{Zhang:2003up} are used to perform an isospin analysis to constrain the CKM angle $\phi_2$ and obtain two solutions with $\phi_2=(93.7\pm10.6)^{\circ}$ being most compatible with other SM based fits to the data. The size of the penguin pollution is consistent with zero: $\Delta \phi_2=(0.0\pm9.6)^{\circ}$. Figure~\ref{fig3} shows the $\phi_2$ can from the isospin analysis.
\section{Summary}
The first observation of $CP$ violation in $\bar{B}^0\to D_{CP}^{(*)}h^0$ decays from a combined analysis of Belle and BaBar dataset is presented. The result is consistent with the value of $\sin(2\phi_1)$ measured in the $b\to c\bar{c}s$ process. Using a similar process,  $B^0\to\bar{D}^{(*)0}h^0$ with $\bar{D}^0\to K_S^0\pi^+\pi^-$, a model-independent time-dependent Dalitz plot analysis is performed and excludes the second $\phi_1$ solution by 5.1 standard deviations. Observation of $B^0\to\psi(2S)\pi^0$ is presented, which will contribute to the $\phi_1$ measurement in future. And finally $\phi_2$ measurement from $B\to\rho\rho$ decays is presented. 
\section*{Acknowledgements}
The author thanks the workshop organizers for
hosting a fruitful and stimulating workshop and
providing excellent hospitality. This research is supported
by the U.S. Department of Energy.


\end{document}